# Electrically driven thermal light emission from individual single-walled carbon nanotubes

David Mann,[1] Y. K. Kato,[1] Anika Kinkhabwala,[1] Eric Pop,[1,2] Jien Cao,[1] Xinran Wang,[1] Li Zhang,[1] Qian Wang,[1] Jing Guo,[3] and Hongjie Dai[1,*]

[1] *Dept. Chemistry and Lab. Adv. Materials, Stanford Univ., Stanford, CA 94305, USA*

[2] *Intel Corp., Santa Clara, CA 95054, USA*

[3] *Department of Electrical and Computer Engineering, Univ. Florida, Gainesville, FL, 32611*

**Light emission from nanostructures exhibits rich quantum effects and has broad applications. Single-walled carbon nanotubes (SWNTs) are one-dimensional (1D) metals or semiconductors, in which large number of electronic states in a narrow range of energies, known as van Hove singularities, can lead to strong spectral transitions.[1, 2] Photoluminescence and electroluminescence involving interband transitions and excitons have been observed in semiconducting SWNTs,[3-9] but are not expected in metallic tubes due to non-radiative relaxations. Here, we show that in the negative differential conductance regime, a suspended quasi-metallic SWNT (QM-SWNT) emits light due to joule-heating, displaying strong peaks in the visible and infrared corresponding to interband transitions. This is a result of thermal light emission in 1D, in stark contrast with featureless blackbody-like emission observed in large bundles of SWNTs or multi-walled nanotubes.[10-12] This allows for probing of the electronic temperature and non-equilibrium hot optical phonons in joule-heated QM-SWNTs.**

We investigated electrically driven thermal light emission of individual QM-SWNTs in both the visible and infrared (infrared) (wavelength λ=500-2100 nm, Supplementary Information), in a wider spectral window than previously explored for electroluminescence of nanotubes. We fabricated suspended and non-suspended SWNT (diameter $d$~2-4nm) devices with tube length $L$ ~ 2-10 μm (Fig.1a, 1c insets), as described previously.[13-15] QM-SWNTs were identified as those exhibiting weak source-drain current ($I_{ds}$) dependence (due to small band-gaps ~ tens of meV [14, 16]) on gate-voltage ($V_{gs}$) with $I_{ds}(max)/I_{ds}(min) < 10$ (at bias $V_{ds}$=10 mV), across the $V_{gs}$ range (Fig. 1a). On substrate, QM-SWNTs show current saturation near 20 μA at high bias, while suspended ones exhibit negative differential conductance (i.e., reduced currents at higher biases) and much lower maximum current <10 μA (Fig.1b) due to joule-heating and electron scattering by hot optical phonons caused by slow heat dissipation in suspended SWNTs.[15]

We observed light emission from suspended QM-SWNTs (in on-state under a high negative $V_{gs}$, with the device kept in Ar) beginning at low $V_{ds}$, with pronounced peaks in the spectra (Fig.1c, 2a&2b). We measured the visible emission characteristics of the suspended and on-substrate sections of several QM-SWNTs (Fig.1c) and observed that the onset of detectable visible light for suspended QM-SWNT devices began as low as $V_{ds}$ = 0.9 V (always in the negative differential conductance region in $I_{ds}$-$V_{ds}$ as the one in Fig.1b), while the visible emission for on-substrate SWNTs was not measurable until $V_{ds} > 5$ V.  For several long (10 μm) suspended QM-SWNTs, we spatially resolved light emission and found that the location of the brightest spot was always near the center (Fig.1d) and remained stationary at various $V_{ds}$ and $V_{gs}$.

We investigated light emission from ten suspended (all in Ar) QM-SWNTs (Fig. 2). All QM-SWNTs exhibited spectral peaks and the peak positions varied (Fig. 2a&2b). In SWNTs, electronic transitions between the van Hove singularities are dipole-allowed (denoted as $E_{nn}$ transitions).[17] We attribute the observed peaks to optical emission (highly polarized along tube axis, Fig. 2b inset) from $E_{11}$ (infrared) and $E_{22}$ (visible or near infrared) transitions (Fig. 2c) of QM-SWNTs. Lorentzian fitting is used to determine the peak locations of $E_{11}$ and $E_{22}$. We find reasonable agreement with simple tight-binding predicted $E_{11}$ and $E_{22}$ values (~1:2 ratio) (Fig. 2d)[18] for QM-SWNTs with $d$~2.8 to 4nm ($d$ was measured by atomic force microscopy over the on-substrate portion of the nanotubes).

To understand the light emission in QM-SWNTs, we note that the negative differential conductance in the $I_{ds}$-$V_{ds}$ of suspended QM-SWNTs is indicative of significant self-heating and electron scattering by hot optical phonons.[15] The slow decay and long lifetimes of optical phonons in suspended SWNTs lead to high non-equilibrium optical phonon population and temperature ($T_{op}$), causing significant electron heating ($T_e$~$T_{op}$) well above the temperature of the SWNT lattice.[15, 19] Analysis of the negative differential conductance region of the $I_{ds}$-$V_{ds}$ curve of a ~2μm suspended QM-SWNT by the hot phonon model[15, 19] leads to an estimated $T_e$ ~ $T_{op}$ ~1200 K at $V_{ds}$ ~ 1.3 V. This heating gives rise to a thermal distribution of electrons and holes with appreciable populations at the van Hove singularities in QM-SWNTs (Fig. 2c). These carriers can then radiatively recombine to produce $E_{11}$ and $E_{22}$ emission peaks, thus producing distinct spectral features rather than featureless blackbody spectrum. Note that excitons may play a role in metallic and QM-SWNTs, but the effect should be smaller in our case than in semiconducting-SWNTs due to large $d$~2-4nm QM-SWNTs used with low exciton binding energies[20] relative to the high $T_e$ involved. The effect is difficult to discern from our spectra with broad peaks caused by significant heating.

This thermal light emission model is consistent with the observed emission photon energy exceeding the bias-voltage injection energy $eV_{ds}$ (emission well above $eV_{ds}$=1.4eV is seen in Fig. 1c). It is also consistent with the drastic difference in light emission between the suspended and on-substrate portions of a SWNT (Fig. 1c), since self-heating of the latter is much lower due to efficient thermal dissipation and optical phonons relaxation into the substrate.[15, 21] In fact, in ambient air without the protection of Ar flow, our suspended

SWNTs breakdown at sustained biases $V_{ds}$~1.5-2V (see Supplementary Fig.S1) as a result of oxidation as their lattice temperature approaches ~800K.[22] The thermal light emission model is further consistent with the fact that light emission is brightest at the center of the suspended QM-SWNTs (Fig.1d) where a parabolic temperature profile peaks.[15] This differs from previous spatially resolved electroluminescence in semiconducting-SWNTs in which emission was observed at the suspended trench edge attributed to impact excitation and exciton recombination[7], and the mobile emission spot seen as a result of ambipolar carrier injection.[5] We carried out theoretical modeling (see Method section) to fit the experimental spectra (Fig. 3a &3c) and extract electron temperatures by spectra fitting in the visible region, and the results are close to the optical phonon temperature ($T_e$~$T_{op}$) derived from the hot phonon model (Fig. 3b left axis).[15, 19] Note that our model is mainly used to fit the exponential emission tail in the visible for extracting electron temperature, not intended to precisely fit the peak positions.

Several features in our spectra are not well understood. First, $E_{11}$:$E_{22}$ ~ 1:1.7-2 has been observed for semiconducting-SWNTs by photoluminescence experiments.[23, 24] In our case of QM-SWNT thermal light emission, in which we do not consider excitonic effects, we expect $E_{11}$:$E_{22}$ ~ 1:2, but deviations from this ratio were observed (Fig.2d). One possible cause is significant heating effect on the nanotube structure and in turn electronic structure. Some of our QM-SWNTs exhibited unexplained peaks (e.g., in red curve of Fig. 2a) between $E_{11}$ and $E_{22}$. Possibilities include phonon assisted transitions, inter-band transitions (e.g., $E_{12}$ for which theoretical work has suggested perpendicular polarization and intensity up to a ~1/5-1/3 of $E_{nn}$ transitions[25]), and possibly emission from states due to defects along the relatively long tubes. These possibilities require further investigations. We calculated the effect of trigonal warping[26] on our spectra and found the effect to be inconsequential in this diameter range (d~2.8-4nm), given the breadth of the measured emission peaks (>100meV).

For the SWNT in Fig. 3c, we have analyzed the peak width (full-width half maximum $\sigma$ ~130 meV) as a function of bias by fitting several spectra ($V_{ds}$= 0.7 to 1.2V). As $V_{ds}$ and thus $T_e$ increases, the emission peak is expected to widen from increased thermal and lifetime broadening effects. Indeed, we observed peak-width change over the bias range (Fig. 3c). The apparent peak-widths correspond to effective lifetimes of $\tau_{TOT}$ ~10 to 14 fs, including all scattering mechanisms (Fig. 3d left axis). By using the calculated $T_{op}$ (and the corresponding Bose-Einstein optical phonon occupation number) from the hot phonon model,[15, 19] we determined an electron-phonon scattering lifetime of $\tau_{e\text{-}op}$ ~ 15 to 18 fs (Fig. 3d right axis), about 50% greater than $\tau_{TOT}$. This suggests that only a portion of $\tau_{TOT}$ is due to electron-phonon scattering, with additional broadening likely due to other mechanisms, such as electron-electron scattering.

Lastly, we carried out light emission measurements of QM-SWNTs as a function of bias $V_{ds}$ and gate-voltage $V_{gs}$ (Fig. 4). At a fixed bias $V_{ds}$, the infrared light emission $\gamma$ (down to 0.57eV) under various $V_{gs}$ scaled exponentially with current or power ($P = I_{ds}V_{ds}$) (Fig. 4a), since increases in the latter caused higher $T_e$. Current modulation by $V_{gs}$ (Fig. 4b, also Fig. 1a) was due to the existence of small band-gaps (~tens of meV) in the QM-SWNTs.[13-16] By simultaneously measuring light emission $\gamma$ and power $P$ vs. $V_{ds}$ and $V_{gs}$ (Fig. 4c and 4d respectively), we observed that the exponential dependence of $\gamma$ on $P$ held across the entire $V_{gs}$ and $V_{ds}$ two-dimensional space.

Figure 4a has striking similarities to the data presented previously.[7] While Chen et al. attribute this to impact excitation and recombination of free carriers and excitons,[7] we rule out impact excitation as cause of light emission in our devices due to the observation of photons of greater energy than the applied field ($E_{photon} > eV_{ds}$) (Fig.1c red curve and Fig.3a). Additionally, we do not expect appreciable light emission from impact excitation as a result of non-radiative relaxation of excited carriers in QM-SWNTs.[20] We have also measured light emission of suspended semiconducting-SWNTs and found that thermal effects also occur in semiconducting tubes (See Supplementary Fig.S2). The data suggests that thermal heating may play a role in other electroluminescence measurements of semiconducting-SWNTs, [3-9] where similar or higher powers than that reported here are dissipated in the devices.

Our measurement gleans the high temperature optoelectronic properties of quasi-metallic SWNTs. By exploiting SWNTs of specific diameters, one can produce thermal light emission peaked at desired wavelength from visible to infrared, which is useful for opto-electronics for telecommunications in 1.3-1.5μm range. While thermal light emission of bulk materials has been extensively studied, our result revealing drastic spectra peaks for SWNTs underlines the importance of examining electronic heating and emission in novel nanomaterials.

**METHODS**

**Experimental details:**

Fabrication of devices, methods for ensuring single tubes, light emission spectra and spatially resolved emission and electrical measurements are described in Supplementary Information.

**Theoretical modelling of thermal light emission spectra of QM-SWNTs:**

We use the tight-binding approximation to calculate the approximate joint density of states $D_J(E) = D(E/2)/2$ (where $E$ is the transition energy and $D$ is the density of states)[18] for a SWNT of a certain diameter $d$, and introduce broadening of the $D_J$ by convolving it with a function $B$ (either Gaussian or Lorentzian),[28]

$$D_J^B(E) = \int_{-\infty}^{+\infty} dE' \cdot D_J(E') \cdot B(E - E', \sigma) \qquad (1)$$

where σ is the broadening width (due to finite lifetime of carriers scattered by phonons and other mechanisms) and used as a fitting parameter. It is important to note that $D_J$ does *not* include the metallic electronic band, since the dipole transition matrix element is zero for that band.[29] As an approximation (without including exciton effects for the large diameter QM-SWNTs used in the current work), we calculated the emission spectrum by[30]

$$S(E) = \frac{1}{\tau(E)} D_J(E) f_0[E_C(k) - F_n]\{1 - f_0[E_V(k) - F_p]\}, \quad (2)$$

where $S(E)$ is the photon count (~light intensity), $E = E_C(k) - E_V(k)$ is the emitted photon energy, $D_J(E)$ is the joint density-of-states, $f_0(E)$ is the Fermi-Dirac distribution at high $T_e$ (resulting from self-heating) and $1/\tau(E)$ is the transition probability. $E_C(k) = -E_V(k)$ if the middle of the bandgap is defined as the energy zero because the conduction and valence bands of SWNTs are symmetric, and we assume that the electron and hole quasi-Fermi levels $F_n \approx F_p \approx 0$ under all experimental gate voltages because the nanotube is quasi-metallic and the gate efficiency factor is small(~0.01)[13-15] (Fermi level modulation by gate voltages involved only leads to a variation of < 20% in the product of the electron and hole Fermi-Dirac population terms in Eq.(2).) The emission rate $\frac{1}{\tau(E)} = \frac{2\pi}{\hbar}\left(\frac{q}{m_0}p_{CV}A\right)^2\left(\frac{2}{3}D_{ph}(E)\right)$ depends on the momentum matrix element $p_{CV}$, the magnitude of the vector potential $A$, and the photon density of states $D_{ph}(E)$, and we assume that it is energy-independent for simplicity.[30] For three-dimensional isotropic (black-body) photons, $D_{ph} \propto \omega^2$ ($E=\hbar\omega$), and $1/\tau(E) \sim p_{CV}^2\omega \sim r_{CV}^2\omega^3$, where the dipole matrix element $r_{CV} = p_{CV}/(im_0\omega)$ and $i$ is the imaginary unit. In a quasi one-dimensional SWNT, the momentum matrix element $p_{CV}$ slightly decreases with energy.[30] The energy dependence of the emission spectrum in Eq. (2) is then dominant by the Fermi-Dirac distribution terms in an exponential manner. From ~1.2eV to 2.0eV in the visible range in which our model fitting is carried out for electron temperature extraction, $1/\tau(E)$ computed using $\sim p_{CV}^2\omega$ varies by a factor of <2, but the product of the Fermi-Dirac distribution terms varies by more than 4 orders of magnitude at $T_e = 1000K$.

Using our model with Gaussian broadening, we obtained excellent fitting of the experimental visible spectra (Fig. 3a) and were able to extract $T_e$ at various $V_{ds}$ (square symbols in Fig. 3b). The emission spectrum in the visible essentially exhibits an exponential decay ($\sim e^{-E/k_B T_e}$ due to the Fermi-Dirac distributions in Eq. 2) into the high energy end, with a superimposing hump at ~1.6eV corresponding to the $E_{22}$ transition (Fig. 3a). Thus the visible spectrum of suspended QM-SWNTs allows for an experimental determination of $T_e$ (~$T_{op}$) for individual SWNTs under joule heating at various $V_{ds}$. Under higher $V_{ds}$, a suspended SWNT exhibits more Joule heating and higher $T_e$ (Fig. 3b left axis), and thus an exponential increase in light emission (see bias dependent spectra in Fig. 3a&3c, and Fig. 3b right axis). Importantly, we found that the extracted $T_e$ (squares in Fig. 3b) from emission spectra agree well with those obtained (Fig. 3b blue line) by fitting $I_{ds}$-$V_{ds}$ curves (Fig. 3b inset) using the hot phonon model.[15, 19]

The emission spectra of suspended QM-SWNTs in the lower energy infrared regime were dominated by the $E_{11}$ peak (Fig.2a&3c). This supports our assumption for the model where the transitions within the metallic band are forbidden,[29] since otherwise an exponentially increasing emission would exist on the lower energy side of the $E_{11}$ peak. It is interesting that thermal light emission spectra gives insights to the magnitudes of the optical transition matrix elements, but this lack of an exponential slope causes difficulty in extracting $T_e$ by spectra analysis in the infrared region. Instead, by using the calculated temperatures from the $I_{ds}$-$V_{ds}$ fits, Lorentzian broadening and allowing only the width $\sigma$ to vary (after fixing the other parameters by fitting one spectrum), we modelled the infrared spectra with excellent agreement with experiment for the SWNT in Fig.3c. This agreement suggests that contrasting SWNT emission spectra in the infrared region with blackbody is not sufficient to exclude the possibility of thermal emission as done in a recent work.[8] The total photon counts $\gamma$ in the infrared (Fig. 3b red line) can be estimated as $\gamma(T_e) \sim Ae^{-E/k_B T_e}$ at various $V_{ds}$ (and in turn various $T_e$) with $E \sim 0.8\text{eV} \sim E_{11}$ (photon energy of the peak in Fig. 3c), in agreement with the measured results (Fig. 3b red symbols).

**FIGURE CAPTIONS**

**Figure 1.** Visible thermal electroluminescence of quasi-metallic SWNTs. **(a)** Current vs. gate-voltage $I_{ds}$-$V_{gs}$ curve for a 2 µm long suspended QM-SWNT device schematically shown in the inset (S: source; D: drain). **(b)** Current vs. bias $I_{ds}$-$V_{ds}$ characteristics of the SWNT in (a) (showing negative difference conductance), together with that of the non-suspended portion of the same tube. **(c)** Visible emission spectra for the suspended and non-suspended portion of the QM-SWNT recorded at low and high bias $V_{ds}$, respectively ($V_{ds}$=1.4V, $I_{ds}$=5 µA, and $V_{ds}$=7 V, $I_{ds}$=21 µA). Inset shows a scanning electron microscopy image (scale bar is 2 µm) of the device with suspended and on-substrate SWNT portions bridging electrodes. **(d)** Visible confocal image of a 10µm suspended QM-SWNT at $V_{ds}$=1.9 V ($I \sim 3$ µA) collected by silicon avalanche photo-detector superimposed on a dark-field optical image (the brightest horizontal lines mark the edge of the electrodes). The strongest light emission is seen at the center of the suspended tube (approximate location traced by the dashed line). Right panel: $\gamma$ line-cut (total photon counts) along the tube length. The resolution of this measurement is nearly diffraction limited (~1µm).

**Figure 2.** Thermal light emission of suspended metallic SWNTs with $E_{11}$ and $E_{22}$ peaks. **(a)** Light emission spectra (scaled for readability) in the infrared for three independent 2 µm long QM-SWNTs (red green and blue) at $V_{gs}$= -20V and $V_{ds}$ = 1.4, 1.1, 1.3 V respectively ($I_{ds}$=6.35, 5.13, 5.95 µA). **(b)** Corresponding light emission spectra for the three tubes in (a) in the visible at $V_{gs}$= -20V and $V_{ds}$ = 1.5, 1.3, 1.5 V ($I_{ds}$=6.15, 4.78, 5.7 µA) respectively. Inset: symbols are measured photon counts for emission polarized at various angles relative to tube axis. Solid line is a $cos^2$ fit. **(c)** Illustration of thermal light emission mechanism in a [*m,n*]=[*24,21*] QM-SWNT (with $d$~3nm, $E_{11}$~0.8eV, $E_{22}$~1.6eV as for the tube in a&b) with red curves). The curve on the blue region corresponds to electron population at various energies calculated by multiplying the density of states (black line) and the Fermi-Dirac distribution (red line) at $T_e$~1200K. The finite populations at the first and second van Hove

singularities are responsible for $E_{11}$ and $E_{22}$ optical emission and depend on $T_e$ and energy exponentially. **(d)** $E_{11}$ vs. $E_{22}$ peak locations from ten suspended QM-SWNT devices determined from their visible and infrared spectra. The red line corresponds to $E_{22} = 2E_{11}$ from the simple tight binding approximation. The peaks were determined using a Lorentzian curve fit. As a result, some of the devices showing asymmetric peaks had some offset due to imperfect fit.

**Figure 3.** Thermal light emission spectra of a 2.9 μm long suspended QM-SWNT compared with theory. **(a)** Visible spectra of a 2.9 μm long suspended QM-SWNT at two biases $V_{ds}$ = 1.2 (blue) and 1.3 V (red) (top-panel). Lower panel shows spectra calculated for a $d$~3nm SWNT with $E_{11}$~0.8eV and $E_{22}$~1.6eV using Eq. 2. Several $d$~3nm SWNTs (e.g., [24,21] and [28,16]) have similar $E_{11}$ and $E_{22}$, making it not possible to uniquely determine [m,n]. **(b)** Left axis: Electron temperature $T_e$ (~ $T_{op}$) vs. bias (blue line) derived by fitting $I_{ds}$-$V_{ds}$ data (symbols in inset) using the hot phonon model.[15, 19] Blue squares are $T_e$ extracted from fitting visible thermal light emission spectra (see Method) in (a) for the two biases. Right axis: Measured γ vs. bias in the infrared region (red symbols) and computed γ (red line) based on $T_e$ derived from the $I_{ds}$-$V_{ds}$ model.[15, 19] **(c)** Thermal light emission spectra in the infrared region for the SWNT at $V_{ds}$ = 0.7 (black), 0.8 (green), 1.0 (blue) and 1.2 V (red) (top panel). Lower panel is calculated spectra using Eq. 2 and $T_e$ at corresponding biases from (b). Note that a more precise theoretical treatment should include any exciton effects in our $d$~2-4nm QM-SWNTs. The exciton binding energies for large diameter QM-SWNTs are unknown but should be smaller than ~80meV theoretically expected for a $d$~0.5nm QM-SWNT.[27] The effect may cause a shift in the emission peak positions, but the shift will be small compared to the large thermal light emission peak width (~130meV). **(d)** Left axis: estimated hot-electron lifetime ($\tau_{TOT}$) at various biases. Right axis: calculated electron- optical phonon scattering time at various biases at corresponding temperatures obtained from $I_{ds}$-$V_{ds}$ analysis.

**Figure 4.** Thermal light emission of suspended QM-SWNTs exhibits exponential dependence on power dissipation in the devices. **(a)** Total power dissipation $P(=I_{ds}V_{ds})$ vs. $V_{gs}$ for a typical QM-SWNT device at $V_{ds}=1$ V (black line, left axis in linear scale) and corresponding emission (total photon count in *log* scale) vs. $V_{gs}$ in the infrared (red line, right axis). The electrical contact resistance is always an order of magnitude less than the suspended SWNT resistance at high bias. Power dissipation due to contact resistance was not excluded from $P$. **(b)** $I_{ds}$ map (color scale bar on top) at various $V_{ds}$ and $V_{gs}$ showing the evolution of $I_{ds}$-$V_{ds}$ versus $V_{gs}$. **(c)** Thermal light emission (γ) map (top: color scale in *log*) in the infrared region at various $V_{ds}$ and $V_{gs}$. **(d)** power dissipation map $P$ (top: color scale bar in linear scale) calculated by $P=I_{ds}V_{ds}$ from (b) at various $V_{ds}$ and $V_{gs}$. The close resemblance between the thermal light emission map in *log* scale and power dissipation map in linear scale strongly suggests that light emission scales exponentially with power and supports the thermal light emission model.

**Acknowledgements**

We thank Professor W.E. Moerner for use of the confocal optical setup. This work was supported in part by MARCO MSD Focus Center and a NSF-NIRT.

H.D., D. M. and Y.K. conceived and designed the experiments. D.M., Y.K., A.K., E.P., J.C, X.W., L.Z., Q.W., J.G. performed the experiments and analyzed data. H.D., D.M. and Y.K. co-wrote the manuscript.

All authors discussed the results and commented on the manuscript.

\* Correspondence and request for materials should be addressed to HD, hdai@stanford.edu

**Competing financial interests**
The authors declare that they have no competing financial interests

Figure 1:

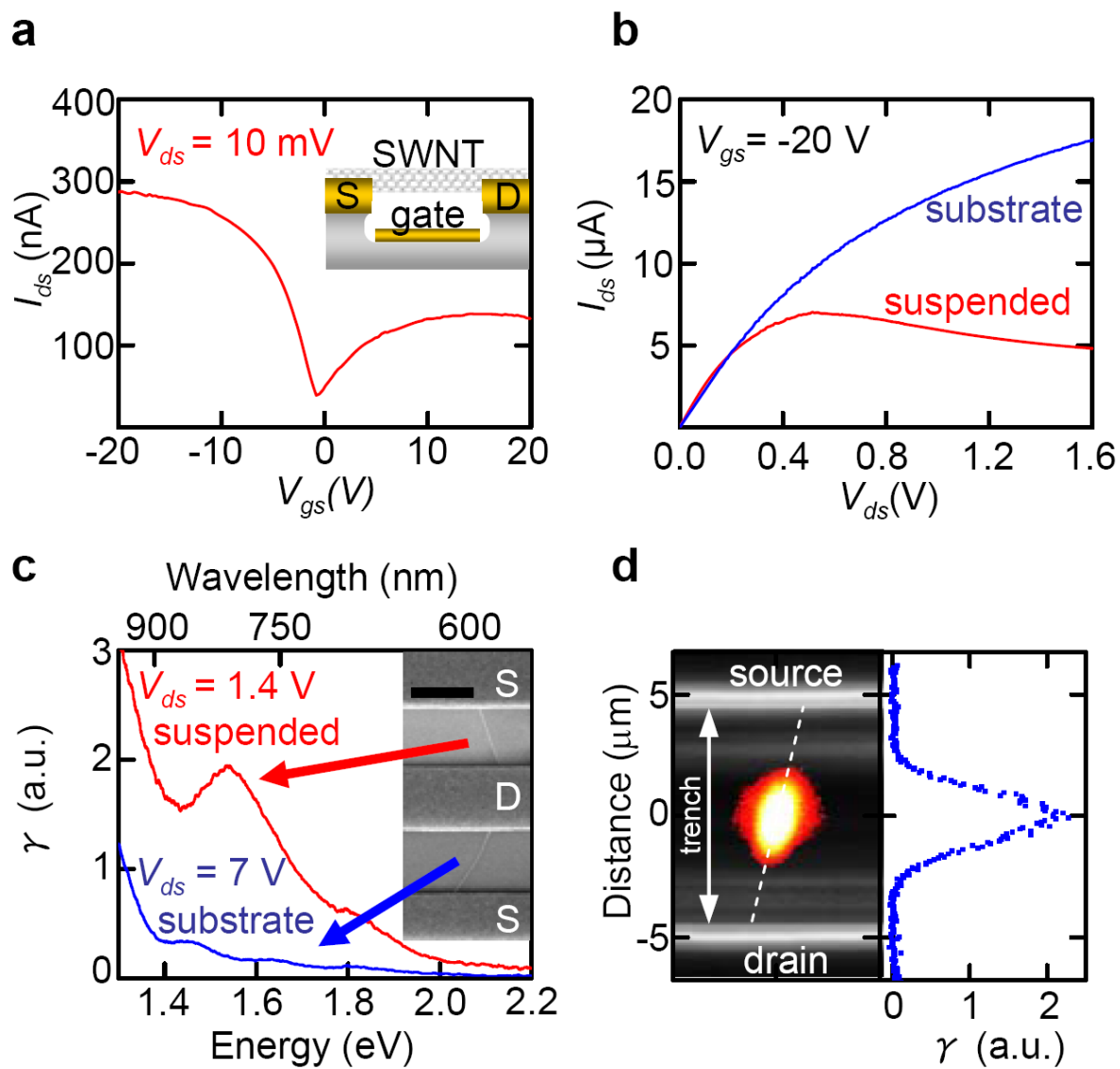

Figure 2:

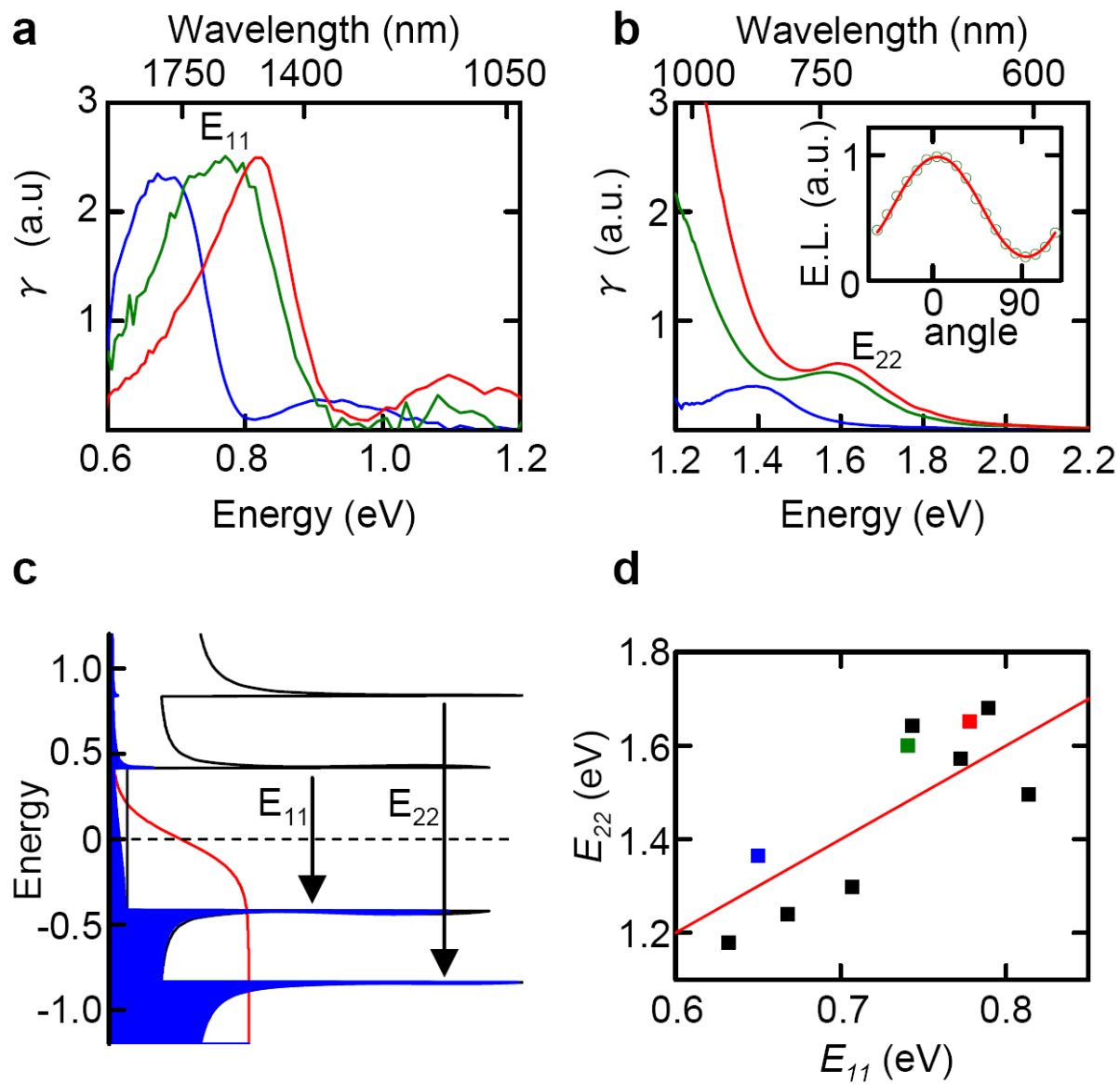

Figure 3:

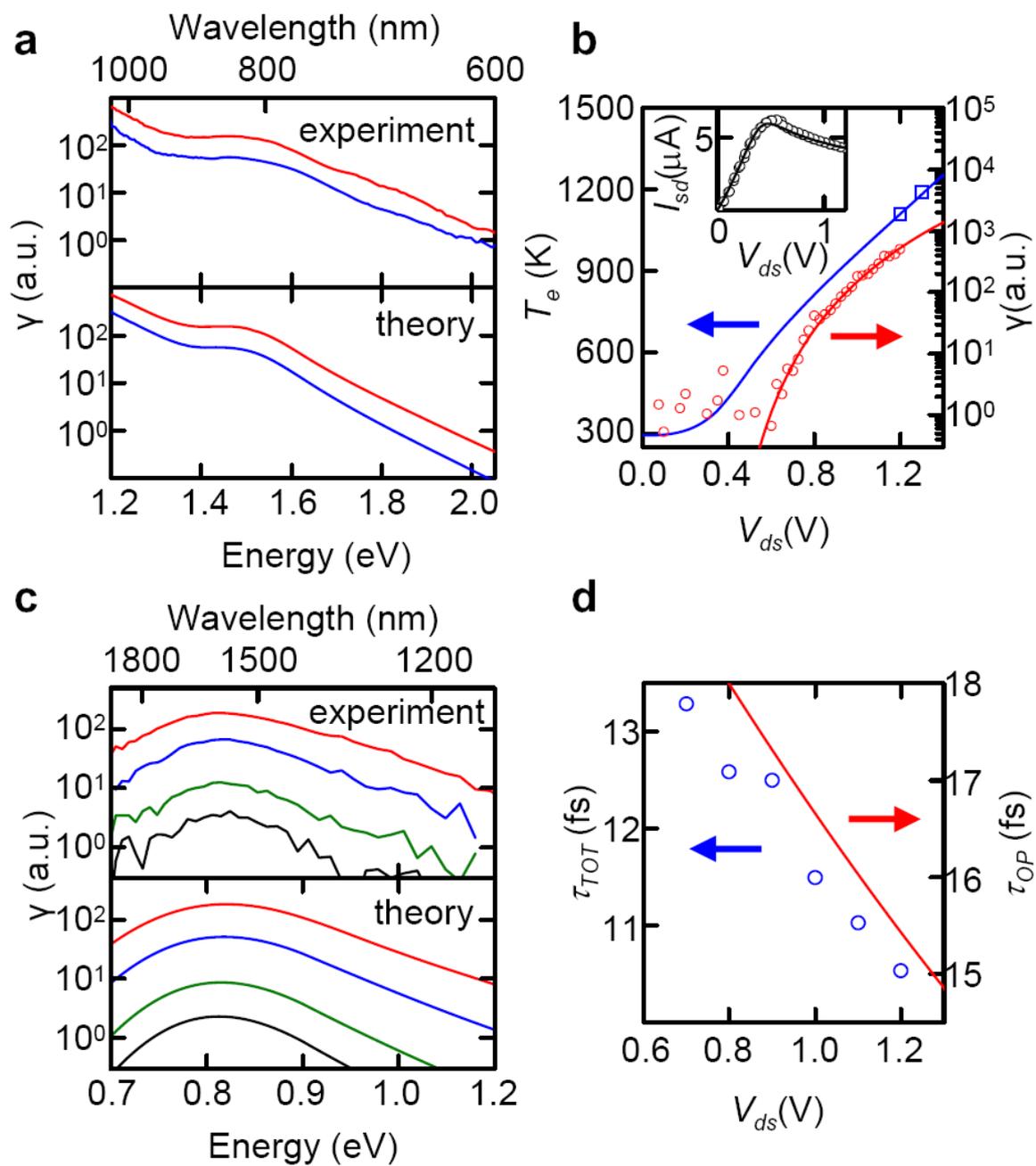

Figure 4:

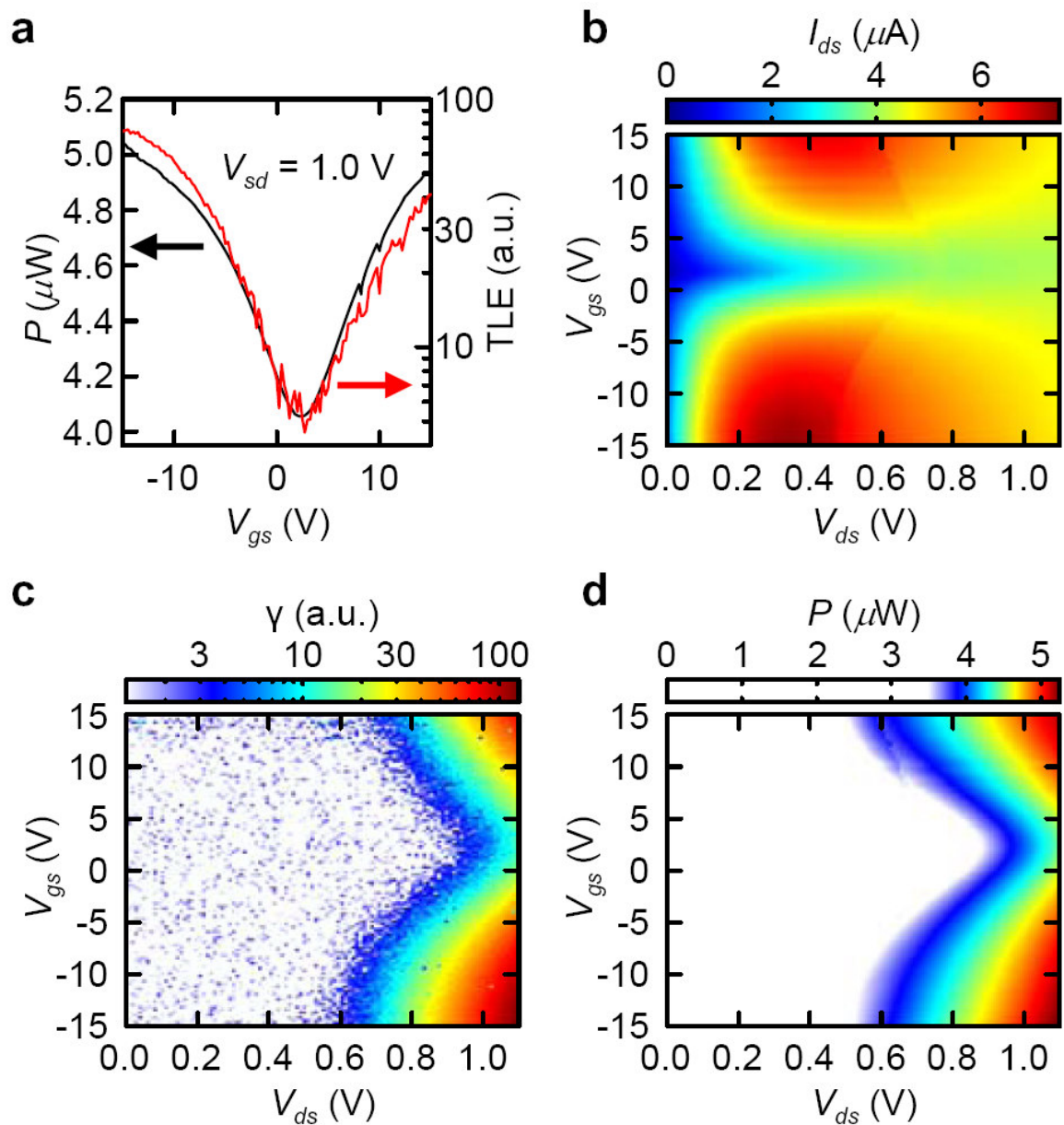

**Supplementary Information.**

**Experimental Details.**

For suspended SWNT devices, we began with photolithographically defined pairs of Pt electrodes spaced 2-10 μm apart (Fig. 1a inset).[1,2] For "suspended/non-suspended" SWNT chips, we began instead with triplets of Pt electrodes with a trench only between the top and middle electrode.[3] (Fig 1c inset) Patterned CVD[4] growth of SWNTs to bridge the electrodes was performed at 825°C.[5]

Devices were bathed in Argon during measurements to prevent breakdown of the SWNTs by thermal oxidation as a result of joule heating at high $V_{ds}$.[6] For thermal light emission (TLE) detection in the visible/near-IR, we used the microscope and detector from a Renishaw micro-raman spectrometer (with 50X objective lens NA = 0.75). A grating with groove density of 1200 mm$^{-1}$ and thermoelectrically-cooled Silicon CCD array (512×512) (detects E > 1.2 eV) were used. For polarization measurements, emitted light passed through a Glan-Thompson polarizer. Typical integration time for spectrum collection was ~60 seconds, though total spectrum collection time was ~20 min due to the high groove density of the grating. For spatially resolved TLE and total emitted light measurements (Fig. 1d), we used a homebuilt scanning confocal microscope (objective lens 80X, NA = 0.8).[7] Light was collected in a Perkin Elmer SPCM-AQR-15 silicon avalanche photo-diode, sensitive down to 1.2 eV. Collection time for each data point was 10ms. We detected on the order of ~100000 counts/second (efficiency of detection systems was roughly 2%, implying roughly ~5x10$^6$ counts/second in the visible/near-IR)

For light detection in the IR, a pair of achromatic lenses were used as an objective (effective NA ~0.6) with a working distance of ~8 mm. The collected light was free-space coupled with an f/# of 7 to an imaging Czerny-Turner spectrometer with a focal length of 300 mm. Detection was done with a liquid-nitrogen-cooled InGaAs linear photodiode array (detects E > 0.58eV). A plane ruled grating blazed at 1700 nm with a groove density of 75 mm$^{-1}$ dispersed the light for obtaining spectra, while the total photon counts were measured by using an aluminum mirror to reflect all wavelengths. The 1/f noise associated with the large dark current of InGaAs photodiodes was minimized by subtracting background counts at ~0.5 Hz, effectively achieving lock-in detection. The spectra were also averaged over all 1024 pixels of the detector to eliminate the effect of response non-uniformity across the array. Typical spectra were collected over 30 min at a resolution of ~17 nm. When a mirror is used in place of the grating, typically around 50,000 counts/s (65 photons/count) were detected.

All spectra were <u>corrected</u> for by using the polarization dependent responsivity of both the visible and IR spectrometers. The responsivity was measured by using a Lindberg-Blue tube furnace (with a large graphite crucible at the center) as a blackbody calibration source.

All electrical measurements were performed at room temperature with an Agilent 4156C semiconductor parameter analyzer and an Agilent 33120A arbitrary function generator (for 0.5 Hz pulse generation)

**Single Tube Devices.**

To ensure that we were measuring single SWNTs in our experiments, we followed a series of standard procedures below.

(1) Catalyst selection: The catalyst we used (Alumina based catalyst) has been shown previously to produce predominately individual tubes of many diameters with very few bundles.[8] We have used variations on this catalyst (varied to produce different diameter ranges) successfully for many electronic studies on individual SWNTs.[2, 9-11]

(2) Growth: We aimed to grow a low number of large diameter tubes by tuning the amount and composition of the catalyst, as well as adjusting the growth temperature and conditions. We aimed for approximately 20% yield, meaning 20% of the devices in each growth succeeded in making an electrical connection. At this low yield, it is unlikely to produce bundles, since the total number of tubes produced is so low. Beyond that, we noted that ~2/3 of the electrically connected devices were semiconductors, which would not be the case if we were measuring bundles.

(3) After growth, we selected devices via electrical probing that were stable, and had peak conduction levels on substrate that did not exceed 25μA (indicative of a single tube) under high biases, and also had clear negative differential conductance (NDC) on the suspended portion (multiple tubes or bundles can lead to kinks and irregularities in the NDC). Further, after the emission measurements were concluded, we took several devices and used electrical breakdown <u>in air</u> (breakdown voltage ~1.5-2.5 V for ~ 2μm long QM-SWNTs) to determine whether there was a single connection or multiple (or bundles) by counting the number of current jumps during breakdown (Fig.S1). Any data that came from a device with multiple connections was discarded, and its characteristics were noted so that we could avoid that type of device in the future ($I_{ds}$-$V_{gs}$, I-V and emission characteristics).

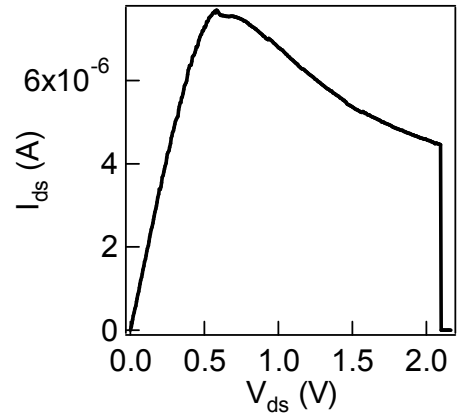

Figure S1: Electrical breakdown of an individual suspended QM-SWNT in air. The sudden current jump to zero at ~ 2V is due to oxidation of the joule heated nanotube.

The electrical breakdown voltages in air (1.5-2.5V) of our QM-SWNT devices are close to but somewhat higher than the upper bound of bias voltages used in our TLE experiments in Ar. In air, during I-V sweeps, SWNTs can survive brief sweeps to higher voltages than constant holding voltages applied across the tubes continuously. This suggests a QM-SWNT can sustain higher temperature for shorter time against oxidation in air.

**TLE of suspended semiconducting SWNTs (S-SWNTs) due to electronic heating.**

Fig.S2a shows TLE spectra of a suspended S-SWNT device taken at two biases ($V_{ds}$) showing a single spectral peak in the IR in the on-state of the device under $V_{gs}$ = -20V. A strong bias dependence of the TLE intensity is seen, similar to the suspended QM-SWNT devices in the main text. Fig.S2b shows a 2-D compilation of many $I_{ds}$-$V_{ds}$ taken from $V_{gs}$ = -20 to 20V for the S-SWNT device. This is a p-channel device, with NDC at $V_{gs}$ = -20 V and no current in the $V_{gs}$ > 0 region.

Fig.S2c shows TLE ($\gamma$) simultaneously collected during $I_{ds}$-$V_{ds}$ map in Fig. S2b, plotted in log scale. Fig.S2d shows the total power $P$ dissipated over device and contacts plotted in linear scale. For any given $V_{gs}$, $\gamma$ scales exponentially with $P$ similar to the device in Figure 4 in the main text for QM-SWNTs. This qualitatively suggests that TLE in suspended S-SWNTs increases as self-heating increases (or as power dissipation increases). In contrast to the QM-SWNT in Figure 4, however, the overall scaling factor for $\gamma$ to $P$ changes a little as a function of $V_{gs}$, as a result of significant contact resistance for S-SWNT devices due to Schottky barriers at the contacts.

To understand the small $P$ to $\gamma$ scaling discrepancy between the QM-SWNT in Figure 4 and the S-SWNT in Figure S2, note first that in both cases the intrinsic contact resistance ($R_c$) is constant across all $V_{gs}$, while the SWNT resistance ($R_{swnt}$) changes with $V_{gs}$. For the QM-SWNT in Figure 4, $R_c$ is very small compared to the $R_{swnt}$ resistance at high bias ($R_c$~15KOhms vs $R_{swnt}$ >150KOhms), which means that for the complete $V_{gs}$ range, the vast majority of the power is dissipated over the SWNT, leading to a simple relationship between total dissipated power and temperature in the SWNT (and thus light emission). In the case of the S-SWNT, $R_c$ is large ($R_c$ ~150KOhms, common for a Pt contacted S-SWNT device). At high bias, $R_{swnt}$ varies from ~ 150KOhms at $V_{gs}$ = -20V to infinity as $V_{gs}$ becomes positive. With $V_{gs}$ = -20V, $R_c$ is on the order of $R_{swnt}$ so a significant part of the total power is dissipated at the contact and does not contribute to SWNT heating. As $V_{gs}$ approaches positive values, $R_{swnt}$>>$R_c$, so almost all of the total dissipated power heats the SWNT and thus contributes to light emission. This leads to the observation of more light emission per input power as $V_{gs}$ sweeps from -20V to +20V in Figure S2c and d.

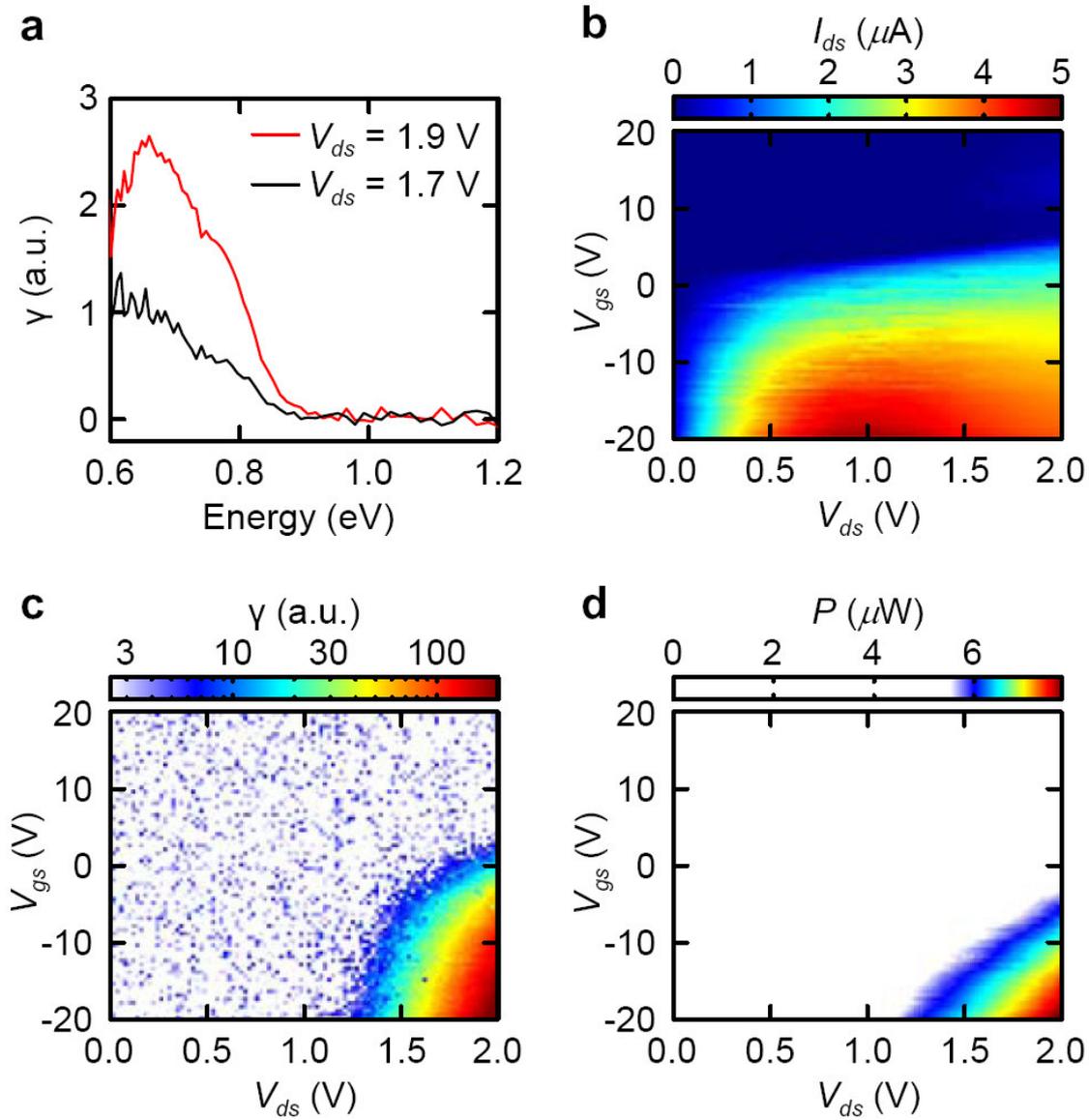

Figure S2: Electrical characteristics and TLE spectra of a semiconducting SWNT device. (a) IR spectra taken at two $V_{ds}$ showing a single spectral peak. (b) A 2-D compilation of many $I$-$V_{ds}$ taken from $V_{gs}$ = -20 to 20. This is a p-channel device, with NDC at $V_{gs}$ = -20 V and no current in the $V_{gs}$ > 0 region. (c) γ collected during I-V map in Figure S1b, plotted in log scale (d) total $P$ dissipated over device and contacts plotted in linear scale. For any given $V_{gs}$ TLE scales exponentially with $P$ similar to the device in Figure 4. In contrast to Figure 4, the overall scaling factor for TLE to $P$ changes a little as a function of $V_{gs}$, as a result of significant contact resistance.